\documentclass{icsm}
\usepackage{graphicx}
\usepackage{bm}
\usepackage{amsmath}
\usepackage{amsfonts}
\usepackage{amssymb}
\usepackage{amsthm}

\usepackage[dvips]{epsfig}

\slacs{.4ex}

\begin{document}
\title{A product formula and combinatorial field theory}
\authori{A. Horzela$^{a}$,  P. Blasiak$^{a,b}$,}
\addressi{$^{a}$H.Niewodnicza\'nski Institute of Nuclear Physics, Polish Academy of Sciences,\\
ul. Eliasza-Radzikowskiego 152, PL 31342 Krakow, Poland\\
$^{b}$Laboratoire de Physique Th\'eorique des Liquides,
Universit\'e Pierre et Marie Curie,\\
Tour 24 -- 2e \'et., 4 Pl.Jussieu, F 75252 Paris Cedex 05, France\\
e-mail: andrzej.horzela@ifj.edu.pl,  blasiak@lptl.jussieu.fr}
\authorii{G. E. H. Duchamp$^{c}$,}
\addressii{$^{c}$Universit\'{e} de Rouen, LIFAR, F 76821 Mont-Saint Aignan Cedex, France\\
e-mail: gduchamp2@free.fr}
\authoriii{K. A. Penson$^{d}$ and A. I. Solomon$^{d,e}$}
\addressiii{$^{d}$Laboratoire de Physique Th\'eorique des Liquides,
Universit\'e Pierre et Marie Curie,\\
Tour 24 -- 2e \'et., 4 Pl.Jussieu, F 75252 Paris Cedex 05, France\\
$^{e}$The Open University, Physics and Astronomy Department,
\\Milton Keynes MK7 6AA, United Kingdom
\\email: penson@lptl.jussieu.fr,  a.i.solomon@open.ac.uk}
\authoriv{}     \addressiv{}
\authorv{}      \addressv{}
\authorvi{}     \addressvi{}
\headauthor{A. Horzela et al.}
\headtitle{A product formula and combinatorial field theory}
\lastevenhead{A. Horzela et al.: A product formula and
combinatorial field theory}
\maketitle
\begin{abstract}
We treat the problem of normally ordering expressions involving
the standard boson operators $a$, $a^{\dagger}$ where
$[a,a^\dag]=1$. We show that a simple product formula for formal
power series --- essentially an extension of the Taylor expansion
--- leads to a double exponential formula which enables a powerful
graphical description of the generating functions of the
combinatorial sequences associated with such functions --- in
essence, a combinatorial field theory. We apply these techniques
to some examples related to specific physical Hamiltonians.
\end{abstract}

\section{Introduction}

The normally ordered form of an expression involving  ladder or,
more generally, field operators is defined as one in which all
annihilation operators are moved to the right using the
appropriate commutation rules. Such expressions play an important
role wherever the Fock space representation is used. In field
theory and in many body quantum mechanics, normal order enters the
formalism through the Wick theorem \cite{[00]}. This enables us to
represent field--theoretical $m$--point Green functions or
statistical $m$--point correlation functions (both defined as the
vacuum mean values of products of the field operators) by
expressions within which all initially taken operators are
contracted. Another advantage of the normally ordered operators is
seen if their matrix elements are calculated in the coherent
states representation \cite{[01]}. If the coherent states are
introduced as eigenstates of the annihilation operators such
matrix elements automatically become functions of the complex
variables and provide us the Fock--Bargmann representation of
quantum mechanical quantities. Quantum optics is one branch of
quantum physics where this approach is a basic tool widely used in
numerous applications.

The intersection between the boson normal ordering problem and
combinatorics was discovered  more than thirty years ago. The
first seminal result \cite{[1]} was that, for both bosons and
fermions, ordering a general string ({\it i.e.}, a product of
nonnegative integer powers) of the canonical creation and
annihilation operators results in an expansion in which the
coefficients are combinatorial numbers called {\it rook numbers}.
The second seminal result \cite{[2]} expressed the boson string
$\left(a^{\dagger}a\right)^n$, where $[a,a^{\dagger}]=1$, as
\begin{equation}
\label{katriel1}
\left(a^{\dagger}a\right)^n=\sum_{k=1}^nS(n,k){a^{{\dagger}k}}a^k\,,
\end{equation}
that is in terms of classical combinatorial numbers known as
\textit{Stirling numbers of the second kind}. Both results,
especially the second one, being the "physicists version" of the
relation
\begin{equation}
\label{katriel2} \left(x\frac{\D}{\D x}\right)^n=
\sum_{k=1}^nS(n,k)x^k \frac{\D^k}{\D x^k}
\end{equation}
known to mathematicians much earlier, have inspired investigations
which lead one to conclude that the relation between the normal
ordering problem and combinatorics is not accidental.
Generalizations of (\ref{katriel1}) to the normally ordered
expressions being more complicated than
$\left(a^{\dagger}a\right)^n$,  \cite{[3]}\,--\,\cite{[10]}, have
led to the introduction of new classes of combinatorial numbers,
whose properties were elucidated using standard methods of
combinatorial analysis --- recurrences, generating functions and
graph representations. Normal ordering for noncanonical operators
has also been investigated, in particular $q$-bosons, leading to
$q$--generalizations of Stirling and Bell numbers \cite{[6]}, \cite{[7]},
\cite{[7a]}. This work has also revealed the relation between
matrix elements taken in the coherent states representation and
the Bell numbers, \cite{[8]}, \cite{[9]}, \cite{[3]}, \cite{[4]}.

As noted above, all these results indicate a fundamental
connection between the general normal ordering problem and
combinatorics. Most of field theoretical calculations employ
operator expressions reduced to the normally ordered form. This
leads one to believe that the methods of combinatorial analysis
will be useful in understanding and solving problems of quantum
physics. Our main goal in this work is to justify this statement.
We shall provide the reader with a general method for constructing
normally ordered expressions and shall explain how to link them to
well-known combinatorial problems. We shall also present analogies
between our methods and those of standard field theory, in
particular Feynman diagrams. Finally, we will illustrate this
approach using examples arising in one mode boson normal ordering.

\section{The product formula}

Let $f(x)=\sum\limits_{n=0}^\infty f_nx^n/n!$ and
$g(x)=\sum\limits_{n=0}^\infty g_nx^n/n!$ be two formal power
series, also called the exponential generating functions (egf) of
sequences $\{f_n\}_{n=0}^\infty $ and $\{g_n\}_{n=0}^\infty$,
respectively. Then
\begin{equation}
\label{prod1} \left.f\left(\lambda\frac{\D}{\D x}\right)g(x)
\right|_{x=0}= \left.g\left(\lambda\frac{\D}{\D x} \right)f(x)
\right|_{x=0}=\sum_{n=0}^\infty f_n\cdot
g_n\,\frac{\lambda^n}{n!}\,,
\end{equation}
which is a straightforward consequence of the elementary relation
\begin{equation}
\label{prod2} \left.\frac{\D^n}{\D
x^n}\left(\frac{x^m}{m!}\right)\right|_{x=0}=\delta_{nm}\,,
\end{equation}
reducing the usual Cauchy product of series to the point--wise
Hadamard product.

If we apply this result to a function $F(\hat{w})$ of an operator
$\hat{w}$ then for any indeterminate $\lambda$
\begin{equation}
\label{prod3} F(\lambda\hat{w})=\left.F\left(\lambda\,\frac{\D}{\D
x}\right)\E^{x\hat{w}}\right|_{x=0}\,.
\end{equation}
On taking the normal form ${\cal N}[\,\cdot\,]$ of the both sides
\begin{equation}
\label{prod4} {\cal
N}[F(\lambda\hat{w})]=\left.F\left(\lambda\frac{\D}{\D
x}\right){\cal N}(\E^{x\hat{w}})\right|_{x=0}\,.
\end{equation}
We emphasize that on the left--hand side above the functional and
operator aspects are mixed while on the right--hand side they are
distinct. The functional aspects are given by a (formal) series in
usual derivatives while the operator aspects are described by an
universal expression --- namely the normally ordered exponential
of $\hat{w}$, in general a \textit{word} $\hat{w}(a,a^{\dagger})$
in terms of the operators $a$, $a^{\dagger}$. This means that
(\ref{prod4}), which we shall call the \textit{product formula},
enables us to reformulate the general normal ordering problem into
a normal ordering of $\E^{x\hat{w}}$. To calculate it explicitly
still remains a non-trivial mathematical task but the problem is
tractable for a large class of physically interesting examples. In
this note we shall consider the cases where
$\hat{w}(a,a^{\dagger})$ is either a product of positive powers of
$a$ and $a^{\dagger}$, or a power of  $a+a^{\dagger}$.

\section{The double\bmth{-}exponential formula}

Using the notation
\begin{equation}
\label{prod5}
\mathcal{N}(\E^{x\hat{w}})\equiv\;:G_{\hat{w}}(x,a,a^\dag):\,,
\end{equation}
where the symbol $:\ :$ denotes that the function
$G_{\hat{w}}(x,a,a^\dag)$ is normally ordered assuming that
$a^\dag$ and $a$ {\em commute}, \cite{[01]}, \cite{[02]}, we can
rewrite (\ref{prod4}) as
\begin{equation}
\label{prod6} \mathcal{N}[F(\lambda\hat{w})]=
\left.F\left(\lambda\frac{\D}{\D x}\right)
:G_{\hat{w}}(x,a,a^\dag):\right|_{x=0}\,.
\end{equation}
For many physical applications the general form of $F$ is given by
\begin{equation}
\label{prod7} F(x)=\exp\left(\sum\limits_{m=1}^\infty
L_m\frac{x^m}{m!}\right)
\end{equation}
and $G$ may also be written in exponential form
\begin{equation}
\label{prod8} :G_{\hat{w}}(x,a,a^\dag):=\;
:\exp\left(\sum\limits_{n=1}^\infty
V_n^{(\hat{w})}(a,a^\dag)\frac{x^n}{n!}\right):\,.
\end{equation}
Substituting (\ref{prod7}) and (\ref{prod8}) in (\ref{prod6}) we
arrive at the so-called \textit{double--exponential formula}
\begin{equation}
\label{prod9}
\mathcal{N}[F(\lambda\hat{w})]=\exp\left(\sum\limits_{m=1}^\infty\frac{L_m}{m!}\,
\lambda^m\frac{\D^m}{\D x^m}\right)\cdot
\left.:\exp\left(\sum\limits_{n=1}^\infty
V_n^{(\hat{w})}(a,a^\dag)\,\frac{x^n}{n!}\right):\,\right|_{x=0}\,,
\end{equation}
which is a convenient starting point for our analysis. Eqn.
(\ref{prod9}) means that using only simple derivations we can find
solution to the normal ordering problem. Thus the function $F$
defines the sequence $\{L_m\}_{m=0}^{\infty}$ while the operator
${\hat w}$ defines the sequence $\{V_n\}_{n=0}^{\infty}$. Examples
of the solutions to the latter problem  are \cite{[10]}
\begin{equation}
\label{prod10}
\begin{array}{lcl}
\hat{w}=a^\dag a\,,&&\disty \mathcal{N}\bigl[\exp(xa^\dag
a)\bigr]=
\,:\exp\left[a^\dag a(\E^x-1)\right]:\,,\\[2pt]
&&V_n^{(\hat{w})}(a,a^\dag)=a^\dag a\,,\\
\noalign{\vskip4pt}
\hat{w}=(a^\dag)^ra\,,&\quad&\mathcal{N}\left[\exp\bigl(x(a^\dag)^ra\bigr)\right]=\\[2pt]
&&\disty=:\exp\left[a^{\dag}a\sum_{n=1}^\infty(a^\dag)^{(r-1)n}(r-1)^n
\frac{\Gamma\left(n+1/(r-1)\right)}{\Gamma\left(1/(r-1)\right)}
\frac{x^n}{n!}\right]\!:\,,\\[9pt]
&&\disty
V_n^{(\hat{w})}(a,a^\dag)=a^{\dag}a(a^\dag)^{(r-1)n}(r-1)^n
\frac{\Gamma\left(n+1/(r-1)\right)}{\Gamma\left(1/(r-1)\right)}\,,\\
\noalign{\vskip4pt}
\hat{w}=a+a^\dag\,,&~~~~&\mathcal{N}\bigl[\E^{x(a+a^\dag)}\bigr]=
:\E^{x^2/2}\E^{x(a+a^\dag)}:\,,\\[2pt]
&&V_1^{(\hat{w})}(a,a^\dag)=a+a^\dag\,,\quad
V_2^{(\hat{w})}(a,a^\dag)=1\,,\\[2pt]
&&V_n^{(\hat{w})}(a,a^\dag)=0\quad\mathrm{for~}\,n>2\,.
\end{array}
\end{equation}

\section{Multivariate Bell polynomials}

Applying the \textit{double exponential formula} requires some
effort. It is easy to see that it works effectively if we deal
with monomials in both exponentials but leads to rather tedious
calculations in more complicated cases. It is more practical to
expand both exponents as formal power series; a general method for
this is the theory of \textit{multivariate Bell polynomials}
\cite{[11]}, \cite{[12]}.

Multivariate  Bell polynomials arose from the  question of
constructing the Taylor--Maclaurin expansion of the composite
function $f(g(x))$. For any $f(x)=\disty\sum_{n=1}^\infty
f_n\frac{x^n}{n!}$ and $g(x)=\disty\sum_{n=1}^\infty
g_n\frac{x^n}{n!}$ given as formal power series one gets
\begin{equation}
\label{bell1} f(g(x))=[f\circ g](x)=\sum_{n=1}^\infty
F_n[f;g]\,\frac{x^n}{n!}\,,
\end{equation}
where
\begin{equation}
\label{bell2}
F_n[f;g]=\sum_{k=1}^nB_{nk}(g_1,g_2,\dots,g_{n-k+1})f_k\,.
\end{equation}
The coefficients $B_{nk}$ are certain polynomials in the Taylor
coefficients $g_i$ called multivariate Bell polynomials, or
sometimes more simply but unprecisely\footnote{The Bell, or
exponential, polynomials are defined as
$B_n(u)=\sum\limits_{k=0}^{n}S(n,k)u^{k}$, {\it i.e.}, as
polynomials which coefficients are the Stirling numbers of the
second kind and $B_{n}(1) = B_{n}$ are the Bell numbers.}, Bell
polynomials.

The multivariate Bell polynomials are closely related to
combinatorial numbers. They satisfy
\begin{equation}
\label{bell3} {B}_{n,k}(g_1,\dots,g_{n-k+1})=
{\sum_{\{\nu_i\}}}^{\prime\prime}\frac{n!}
{\prod\limits_{j=1}^{n}\left[\nu_j!\,(j!)^{\nu_j}\right]}\,
g_1^{\nu_1}g_2^{\nu_2}\dots g_{n-k+1}^{\nu_{n-k+1}}\,,
\end{equation}
where the summation ${\sum\limits_{\{\nu_i\}}}^{\prime\prime}$ is
over all possible non-negative $\{\nu_i\}$ which are partitions of
an integer $n$ into sum of $k$ integers, \textit{e.i.},
\begin{equation}
\label{bell4} \sum_{j=1}^n j\nu_j=n\,,\quad \sum_{j=1}^n\nu_j=k\,.
\end{equation}
From (\ref{bell3}) and (\ref{bell4}) one may show that the
multivariate Bell polynomials satisfy, for $a$ and $b$ arbitrary
constants, the homogeneity relation
\begin{equation}
\label{bell5-1} B_{n,k}(ab^1g_1,ab^2g_2,\dots,ab^{n-k+1}g_{n-k+1})
=a^k b^n\,B_{n,k}(g_1,g_2...,g_{n-k+1})\,.
\end{equation}
Recalling (\ref{prod10}) one expects that the multivariate Bell
polynomials of especial use to us are those from exponential
generating functions
\begin{equation}
\label{bell1-1}
\begin{array}{rcl}
\exp\bigl(g(x)\bigr)&=&\disty\sum_{n=1}^\infty
Y_n[g]\frac{x^n}{n!}\,,\\[6pt]
Y_n[g]&=&\disty\sum_{k=1}^n B_{nk}(g_1,g_2,\dots,g_{n-k+1})\,,
\end{array}
\end{equation}
obtained from (\ref{bell1}) and (\ref{bell2}) for $f_k=1$,
$k=1,2,\dots$.

It may also be seen from (\ref{bell3}) and (\ref{bell4}) that
$B_{n,k}(1,\dots,1)$ are the Stirling numbers of the second kind.
In such a case (\ref{bell2}) gives
\begin{equation}
\label{bell4-1} \exp\bigl(u(\exp{x}-1)\bigr)=
\sum_{n=1}^{\infty}\left(\sum_{k=1}^{n}S(n,k)u^k\right)
\frac{x^n}{n!}\,,
\end{equation}
and
\begin{equation}
\label{bell4-2}
\exp\bigl(\exp{x}-1)\bigr)=\sum_{n=1}^{\infty}B_{n}\frac{x^n}{n!}\,,
\end{equation}
with $B_{n}$ denoting the Bell numbers.

The second useful case is
$\disty\exp{\left(g_1x+g_M\,\frac{x^M}{M!}\right)}$,
\textit{i.e.}, the case when only two g's do not vanish. We have
\begin{equation}
\label{bell4-3}
\begin{array}{c}
\disty\exp\left(g_1x+g_M\,\frac{x^M}{M!}\right)=
\sum_{n=0}^{\infty}H_n^{(M)}(g_1,g_{M})\frac{x^n}{n!}\,,\\[6pt]
H_n^{(M)}(g_1,g_{M})=n!\disty\sum_{r=0}^{[n/M]}
\frac{g_1^{n-Mr}g_{M}^r}{(n-Mr)!r!(M!)^r}\,,
\end{array}
\end{equation}
where $H_n^{(M)}(g_1,g_{M})$ are called the two variable
Hermite--Kamp\'e de F\'eriet polynomials. They are generalizations
of the standard Hermite polynomials, \cite{[12a]}.

Another important property is the {\it inversion formula}. This
states that the following two expressions
\begin{equation}
\label{bell5} Y_n\left[g\right]=\sum_{k=1}^n
B_{n,k}(g_1,\dots,g_{n-k+1})
\end{equation}
and
\begin{equation}
\label{bell6} g_n=\sum_{j=1}^n(-1)^{j}(j-1)!{B}_{n,j}
\bigl(Y_1[g],\dots,Y_{n-j+1}[g]\bigr)\,,
\end{equation}
are inverse to each other if this notion is understood in the
following way: an arbitrary series $\{Y_n[g]\}_{n=1}^\infty$ can
be obtained in the form (\ref{bell5}) if we choose the series
$\{g_n\}_{n=1}^\infty$ only as in (\ref{bell6}). It means that we
are really able to change the double exponential formula into
power series and \textit{vice versa}. Many other properties of the
multivariate Bell polynomials are also known, together with their
explicit forms for some basic (elementary) functions.

\section{The coherent states representation and the analogy to field theory}

As mentioned in the Introduction using the coherent states
representation of the normally ordered strings of $a$ and
$a^{\dagger}$ allows us to dispense with operators and deal with
functions of a complex variable $z$ and its conjugate $z^{*}$.
Taking the coherent states mean value of the \textit{double
exponential formula} we get
\begin{equation}
\label{coh1}
\bigl<z\bigr|\mathcal{N}[F(\lambda\hat{w})]\bigr|z\bigr>=
\left.\exp\biggl(\sum_{m=1}^\infty\frac{L_m}{m!}\,\lambda^m
\frac{\D^m}{\D x^m}\biggr)\cdot \exp\biggl(\sum_{n=1}^\infty
V_n^{(\hat{w})}(z,z^*)\frac{x^n}{n!}\biggr)\right|_{x=0}\,,
\end{equation}
which, for the case $z=1$ and for
$V_n^{(\hat{w})}:=V_n^{(\hat{w})}(1,1)$, becomes
\begin{equation}
\label{coh2}
\bigl<1\bigl|\mathcal{N}[F(\lambda\hat{w})]\bigr|1\bigr>=
\left.\exp\biggl(\sum_{m=1}^\infty\frac{L_m}{m!}\,
\lambda^m\frac{\D^m}{\D x^m}\biggr)\cdot
\exp\biggl(\sum_{n=1}^\infty n^{(\hat {w})}
\frac{x^n}{n!}\biggr)\right|_{x=0}\,,
\end{equation}
identical with the \textit{counting formula} of \cite{[13]}, used
there in order to enumerate the Feynman--type diagrams in
zero--dimensional analogues of the field theoretical models.

The field theoretical analogy may be pushed further on recalling
the functional formalism of the field theory, \cite{[14]}. A basic
quantity which defines any field theory model is the generating
functional of the Green functions. Physically it is interpreted as
the vacuum--vacuum transition amplitude of the time--ordered
exponential of the quantum field operator in the Heisenberg
picture ${\hat\phi}_{\rm H}(x)$ coupled to an external current
$J(x)$
\begin{equation}
\label{coh2-1} G(J)=\left(0\left|T\exp\left(\I\int\D^{D}x
{\hat\phi}_{\rm H}(x)J(x)\right)\right|0\right)
\end{equation}
from which the $m$--point Green functions are got as the $m$--th
functional derivatives with respect to $J$. Passing to the
interaction picture one obtains
\begin{equation}
\label{coh2-2} G(J)=\bigl<0\bigr|T\exp\left(\I\int\D^{D}x \left(
\mathcal{S}_{\mathrm{int}}({\hat\phi}_{\mathrm{I}}(x))+
{\hat\phi}_{\mathrm{I}}(x)J(x)\right)\right)\bigr|0\bigr>,
\end{equation}
where ${\cal S}_{\mathrm{int}}$ is an interaction Lagrangian
density. Rewriting $G(J)$ in terms of a functional integral
\begin{equation}
\label{functint}
G(J)=\int[\D\phi]\exp\left(\I\left(S_0+S_{\mathrm{int}}+{\phi}J\right)\right),
\end{equation}
with $S_0=\int\D^{D}x{\cal L}_{0}$ and $S_{\rm
int}=\int\D^{D}x{\cal L}_{\rm int}$ denoting free (bilinear) and
interaction action functionals for the field $\phi(x)$, and $\phi
J=\int\D^{D}x{\phi}(x)J(x)$, respectively,  one notes that $G(J)$
is an analogue of the partition function of statistical mechanics.
Another expressions for the generating functional of the Green
functions are those  given using functional differential operators
\begin{equation}
\label{coh3}
\begin{array}{rcl}
G(J)&=&\disty\exp\left(\frac12\,\frac{\delta}{\delta\phi}\,
\Delta\frac{\delta}{\delta\phi}\right)
\exp\left(\I\left[S_{\mathrm{int}}(\phi)+\phi J\right]\right)
\Big|_{\phi=0}=\\[12pt]
&=&\disty\exp\left[\I S_{\mathrm{int}}\left(-\I\frac{\delta}
{\delta J}\right)\right]\cdot \exp\left[-\frac12\,J\Delta
J\right],
\end{array}
\end{equation}
where $\Delta=\Delta(x,y)$ is the causal Green function of the
free field equation generated by a free action, $S_{\rm int}$ is,
as previously, an interaction action and we use abbreviations
\begin{equation}
\label{coh4}
\begin{array}{rcl}
\disty\frac{\delta}{\delta\phi}\,\Delta\,\frac{\delta}{\delta\phi}&=&
\disty\int\D^{D}x\,\D^{D}y\,\frac{\delta}{\delta\phi(x)}\,\Delta(x,y)
\frac{\delta}{\delta\phi(y)}\,,\\[9pt]
J\Delta J&=&\disty \int\D^{D}x\,\D^{D}yJ(x)\Delta(x,y)J(y)\,,\\[9pt]
J\phi& =& \disty\int\D^{D}xJ(x)\phi(x)\,.
\end{array}
\end{equation}
Equivalence of both formulas of (\ref{coh3}) comes from the
identity
\begin{equation}
\label{coh5} \left.D\left(-\I\frac{\delta}{\delta J}\right)T(J)=
T\left(-\I\frac{\delta}{\delta\phi}\right)
\left[D(\phi)\exp\bigl(\I\phi J\bigr)\right]\right|_{\phi=0}
\end{equation}
satisfied by arbitrary functionals $D$ and $T$ having formal
Taylor expansions around zero \cite{[15]}. Here we remark that
while in the derivation of the first equality of (\ref{coh3}) the
normal ordering is extensively used, the second equality, as well
as (\ref{coh5}), may be obtained by manipulating the functional
integral representation of the Green functions generating
functional. Although much simpler that the standard advanced field
theoretical methods, our approach give essentially the same
formulae, which are directly applicable to the time evolution
operator or to the {\it partition function integrand}
\cite{[15-1]}.  This we consider as  a strong argument in favour
of the present method.

\section{The counting formula, graphs and combinatorial field theory}

In the effort  to understand the meaning of perturbation
expansions in quantum physics, both in quantum mechanics and in
quantum field theory, it is important to know their large order
behaviour. Solving the problem for the coupling constant
perturbation series one finds that the number of the Feynman
diagrams, contributing order by order to the perturbation series
coefficients, grows factorially, \cite{[15a]}, \cite{[15b]},
\cite{[15c]}.  The factorial growth occurs because of the
combinatorial reasons which (at least qualitatively and for the
large orders  asymptotics) explain why such a behaviour does not
essentially depend on details of the model if one disregards the
problem of \textit{renormalons}\footnote{Renormalons are a
peculiar subclass of the Feynman diagrams which renormalized
values grow factorially with the order of the graph. Renormalons
appear in renormalizable models and are known to influence
perturbation expansions of the nonabelian gauge theories.}.
Enumeration of diagrams for more complicated models is echoed by
that of their zero--dimensional analogues, \cite{[15d]},
\cite{[16]}. Taking all diagrams of a given order to have equal
values shows that this is the number of graphs which determines
the divergence of the (renormalized) perturbation series.  Being
able to estimate these numbers order by order we may investigate a
perturbation series, estimate its asymptotic character and
investigate resummation methods. The idea is to search for
generating functions of given sequences of (combinatorial)
numbers, {\it i.e.,} to look for functions which enumerate the
same weighted objects as the initial formal series do. If found in
analytic form such functions may be treated as generalized sums of
formal, {\it i.e.,} divergent and so analytically meaningless,
series. We shall call such an approach a "combinatorial field
theory" emphasizing that the notion should be understood in a two
senses. The first sense is to obtain a generating function for the
perturbation series, using any of the standard methods. The
second, or inverse sense, is to construct interaction which leads
to a postulated set of combinatorial numbers --- following the
statement of reference \cite{[13]} that \textit{given a sequence
of numbers $\{a_{n}\}$, it is always possible to find a set of
Feynman rules that reproduce that sequence}.

The zero--dimensional analogue of (\ref{coh3}) taken for $J=0$
({\it i.e.}, the generating functional for the \textit{bubble}
Feynman graphs in the field--theoretical jargon) is a particular
case of the right hand side of (\ref{coh2}). So the latter formula
may be considered as a recipe which counts the Feynman--like
graphs in a model for which the potential is
$\disty\sum_{n=1}^\infty V_n\frac{x^n}{n!}$ and for which we deal
with "multilegged propagators" of strenghts $L_{m}$, \cite{[16]},
\cite{[16a]}. Using (\ref{bell4-1}) and (\ref{bell4-2}) we have
for $g$'s related both to $\{L_{m}\}$ and $\{V_{n}\}$
\begin{equation}
\label{feynman1} \E^{ug(x)}=\sum_{n=0}^{\infty}\frac{x^n}{n!}
\sum_{k=1}^nu^k{Y}_{nk}({[g]})=
\sum_{n=0}^\infty\frac{x^n}{n!}{Y}_n({[g]},u)\,,
\end{equation}
and for (\ref{coh2}) we get
\begin{equation}
\label{feynman2}
\bigl<1\bigr|\mathcal{N}[F(\lambda\hat{w})]\bigl|1\bigr>=
Z({L},{V},\lambda)=\sum_{n=0}^\infty A_n({L},{V})
\frac{\lambda^n}{n!}\,,
\end{equation}
using a notation
\begin{equation}
\label{feynman3}
A_n({L},{V})=Y_n([L])\cdot Y_n([V])\,.
\end{equation}

The standard example which illustrates how the counting formula
works is the the zero--dimensional analogue of the anharmonic
oscillator or $\phi^4$ model. It gives
\begin{equation}
\label{feynman4} \exp\left(\frac{a}{2!}\,\frac{\D^2}{\D
x^2}\right)\cdot\exp\left(\frac{-gx^4}{4!}\right)\Big|_{x=0}=
\frac1{\sqrt{\pi}}\sum_{n=0}^{\infty}\frac{\Gamma(2n+\frac12)}{n!}
\left(\frac{-ag}{4!}\right)^n\,.
\end{equation}
This is a consequence of
\begin{equation}
\label{feynman4-1} \exp\left(y\frac{\D^2}{\D x^2}\right)x^n=
H_n(x,y)\,,
\end{equation}
where $H_n(x,y)=H^{(2)}_n(x,y)$ are the Hermite--Kamp\'e de
F\'eriet polynomials, quoted in the Sect.4, (\ref{bell4-3}). For
$x=0$ they take nonvanishing values only for even $n$
\begin{equation}
\label{feynman4-2}
H^{(2)}_{2k}(0,y)=\frac{(2k)!}{k!}\,\frac{y^k}{2^k}\,.
\end{equation}
The expansion in (\ref{feynman4}) may be equally considered either
as the expansion in the number of vertices, {\it i.e.}, in $g^{n}$
or as the expansion in the number of lines, {\it i.e.}, in $a^n$.
In (\ref{feynman2}) the expansion parameter is $\lambda$ which
counts derivatives and, in a graph representation, produces lines
or "propagators". We shall adopt this convention in what follows;
however, each approach may be translated into the other using
(\ref{coh5}).

The series at the right hand side of (\ref{feynman4}) is divergent
which poses the question as to its meaning. The answer is given if
one notices that the same series is got when one calculates a
formal power series expansion in $g$ for the integral
\begin{equation}
\label{feynman5} I(a,g)=\frac1{\sqrt{\pi a^{1/2}}}
\int_{-\infty}^{\infty}\D x\exp\left(-\frac{x^2}{\sqrt{a}}-
\frac{gx^4}{4!}\right).
\end{equation}
The integral above does not define a function analytical for $g=0$
because it does not exist for $g < 0$. Being well defined for
$g>0$ it admits an analytic continuation to
$g\in\{C^2\setminus[0,-\infty)\}$ which is given by
\begin{equation}
\label{feynman6}
I(a,g)=\frac2{\sqrt{2\pi}}\left(\frac3{ag}\right)^{1/2}
\exp\left(\frac3{ag}\right)K_{1/4}\left(\frac3{ag}\right),
\end{equation}
where $K$ is the Macdonald function, {\it i.e.}, a Bessel function
of the imaginary argument. This allows to give a meaning to the
series in (\ref{feynman4}) --- it is an asymptotic expansion of
(\ref{feynman6}) for $g\rightarrow0$,
$g\in\{C^2\setminus[0,-\infty)\}$. So the series under
consideration may be treated as summable in a generalized sense
(nonuniquely, \textit{modulo} functions obeying zeroth asymptotic
expansions) to the properly defined function.

As explained in the Sect.4 we are able to find
$\{A_{n}\}_{n=0}^{\infty}$ either constructing the multivariate
Bell polynomials for sequences $\{L\}_{n=0}^{\infty}$ and
$\{V\}_{n=0}^{\infty}$ or by identifying factors in the
exponential formula with known exponential generating functions.
Formula (\ref{feynman2}) also gives $A_n$ a combinatorial
interpretation. Namely, the number $A_n$ is obtained as the number
of all, connected and disconnected, graphs with the same number
$n$ of labelled lines. The Feynman--like graphs representation is
however different from the conventional one of field theory. For
the first thing, we classify graphs using the number of lines, not
vertices. This does not matter for the simplest examples having
monomials in both exponentials, but for more complicated normally
ordered expressions our method leads to an alternative
description. We refer the reader to \cite{[10]} and \cite{[17]}
for details; here we list only the rules of the graph construction
and define those multiplicity factors which are necessary in order
to understand pictorial illustrations of examples in the next
section. They are:

\noindent
--- a line starts from a white dot, the \textit{origin}, and ends
at a black dot, the \textit{vertex},

\noindent
--- we associate strengths $V_k$ with each vertex receiving $k$
lines and multipliers $L_m$ with a white dot which is the origin of
$m$ lines,

\noindent
--- to count  such graphs we calculate their multiplicity due to the
labelling of lines and the factors $L_m$ and $V_k$.

\section{Examples}

We illustrate our approach using as examples the $\hat{w}=a^\dag
a$ and $\hat{w}=a^\dag+a$. We first consider the normal ordering
problem
\begin{equation}
\label{ex11} \mathcal{N}\left[\exp\left(\lambda a^\dag
a+\frac{\lambda^M}{M!}(a^\dag a)^M\right)\right],
\end{equation}
related to Hamiltonians $\disty\lambda a^\dag a+
\frac{\lambda^M}{M!}\,(a^\dag a)^M$, in particular for $M=2$ to
the Kerr--type Hamiltonian
\begin{equation}
\label{kerr} \mathcal{H}=\lambda a^\dag a
\left(1+\frac{\lambda}2\,a^\dag a\right).
\end{equation}
In our previously introduced notation
\begin{equation}
\label{ex12} L_1=1\,,\quad L_M=1\quad\mbox{for~some~}\,M>1\,,
\quad\mathrm{and}\quad L_m=0\quad\mathrm{otherwise}\,,
\end{equation}
and
\begin{equation}
\label{ex13} F(x)=\exp\left(x+\frac{x^M}{M!}\right);\qquad
V_n^{(\hat{w})}=1\quad{\rm for}\quad n=1,2,\ldots\,.
\end{equation}
In order to get the Taylor--Maclaurin expansion of $F(x)$ we
recall (\ref{bell4-3}) and expand (\ref{ex13}) in terms of the two
variable Hermite--Kamp\'e de F\'eriet polynomials $H_n^{(M)}(x,y)$
\begin{equation}
\label{ex15} F(x)=\exp\left(x+\frac{x^M}{M!}\right)=
\sum_{n=0}^\infty H_n^{(M)}(1,\sfrac{1}{M!})\,\frac{x^n}{n!}\,.
\end{equation}
From (\ref{prod10}), (\ref{feynman2}) and (\ref{feynman3}) this
yields
\begin{equation}
\label{ex15a} A_n=H_n^{(M)}\left(1,\sfrac{1}{M!}\right)\cdot
B_n\,,
\end{equation}
where $B_{n}$ are the Bell numbers. Combinatorially $B_n$ count
all the partitions of an $n$--set and $H_n^{(M)}(1,\frac{1}{M!})$
count partitions of an $n$--set into singletons and $M$--tons. For
$M=2$
\begin{equation}
\label{ex17}
H_n^{(2)}(1,\sfrac{1}{2})=\left(\frac{\I}{\sqrt{2}}\right)^n
H_n\left(-\frac{\I}{\sqrt{2}}\right)=1,\,2,\,4,\,10,\,26,\,76,\,232,\,\dots\,.
\end{equation}
are the \textit{involution} numbers expressible using Hermite
polynomials $H_n(x)$ and the initial terms of $A_n$ are: 1, 4, 20,
150, 1352, 15428, \dots. For $M=3$ the coefficients $A_n$ are: 1,
2, 10, 75, 527, 6293, \dots, {\em etc}. In both cases the series
diverge, as may  seen from the $n\rightarrow\infty$ asymptotics,
\cite{[18]}
\begin{equation}
\label{ex17-1} B_n\sim n!\frac{\exp\bigl(\exp r(n)-1\bigr)}
{[r(n)]^{n+1}\sqrt{2\pi\exp{r(n)}}}\,,
\end{equation}
where $r(n)\sim \log{n}-\log(\log{n})$, applying d'Alembert
criterion to $A_{n}/n!$.

\bfg[ht]                     
\bc                         
\resizebox{8.5cm}{!}{\includegraphics{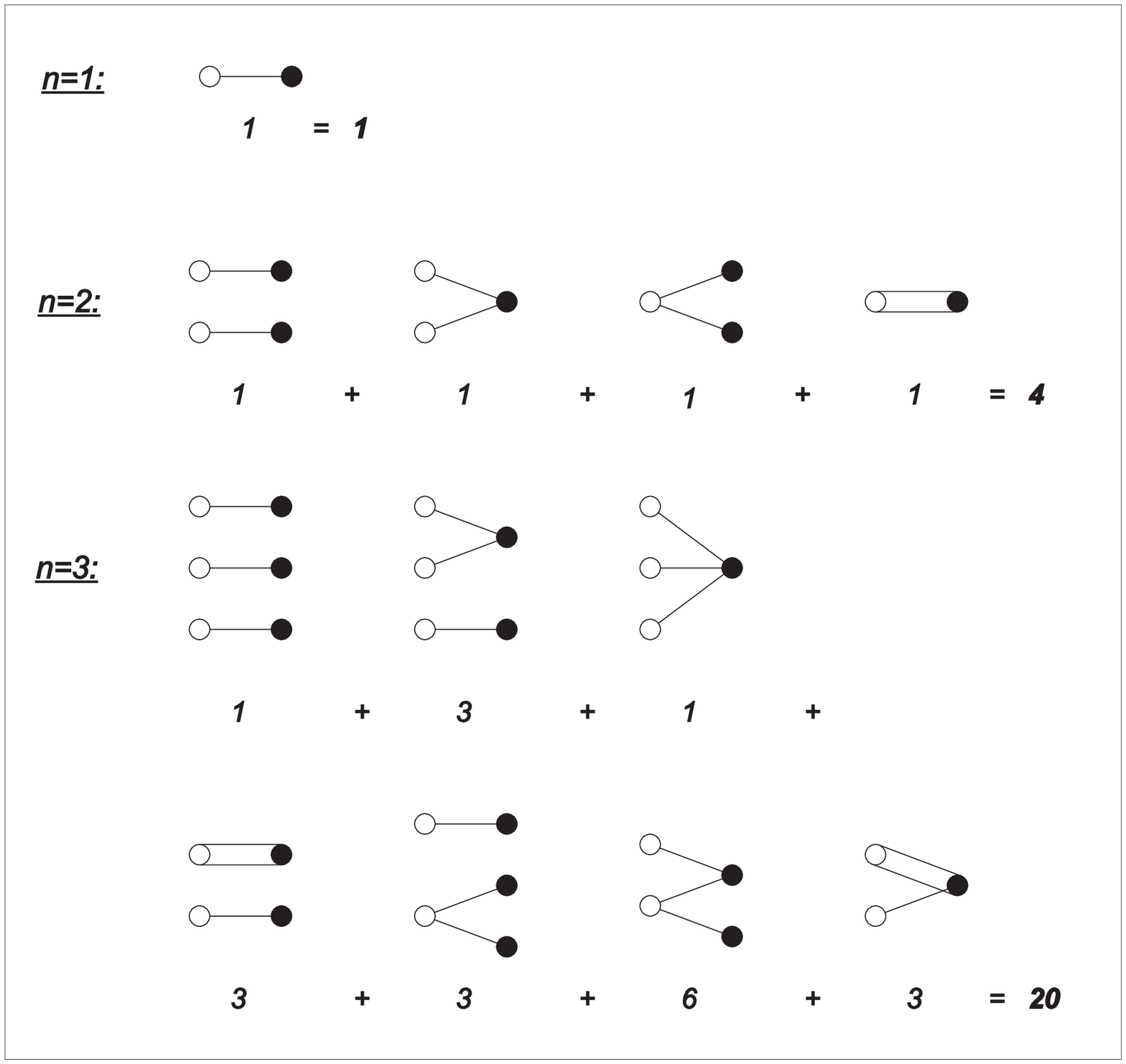}}
\ec                         
\vspace{-7mm} \caption{Lowest order Feynman--type graphs for the
Kerr--type Hamiltonian with $n=1,2,3$ lines. The number below each
graph is its multiplicity.}
\efg                        

We now obtain the generating function of the
sequence$\{A_n/n!\}_{n=0}^{\infty}$. First consider the
\textit{partition function integrand} corresponding to the
Hamiltonian related to (\ref{ex11}) for the particular value
$z=1$:
\begin{equation}
\label{ex18}
\begin{array}{l}
\disty\Bigl<1\Bigr|\mathcal{N}\Bigl[\exp\left(-\beta\left(\lambda
a^{\dagger}a+\sfrac{\lambda^M}{M!}(a^\dag
a)^M\right)\right)\Bigr]\Bigr|1\Bigr>=\\[6pt]
\disty\hspace{20mm}=\sum_{n=0}^{\infty}H_n^{(M)}\!
\left(-\beta,\sfrac{-\beta}{M!}\right)\!\cdot\!B_n\frac{\lambda^n}{n!}=\\[6pt]
\disty\hspace{20mm}=\frac1{\E}\int_0^{\infty}\D x
\sum_{n=0}^{\infty}H_n^{(M)}\!\left(-\beta,\sfrac{-\beta}{M!}\right)
\frac{(x\lambda)^n}{n!}
\left(\sum_{k=0}^{\infty}\frac{\delta(x-k)}{k!}\right)=\\[12pt]
\disty\hspace{20mm}=\frac1{\E}\int_{0}^{\infty}\D x
\exp\left[-\beta\left(x\lambda+\frac{(x\lambda)^M}{M!}\right)\right]
\left(\sum_{k=0}^{\infty}\frac{\delta(x-k)}{k!}\right)=\\[6pt]
\disty\hspace{20mm}=\frac1{\E}\sum_{k=0}^{\infty}\frac1{k!}\,
\exp\left[-\beta\left(k\lambda+\frac{(k\lambda)^M}{M!}\right)\right],
\end{array}
\end{equation}
where we have expressed the Bell numbers as the Stieltjes moments
of the so-called \textit{Dirac comb}, equivalent to the
representation of the Bell numbers by the Dobi\'nski formula.
Applying formal manipulations we obtain a series which is
convergent for positive $\beta$ and which is identical with the
expression which we would get if calculated
$\Bigl<1\bigr|\mathcal{N}\Bigl[\exp\left(-\beta(\lambda a^\dag
a+\frac{\lambda^M}{M!}(a^\dag a)^M)\right)\bigr]\bigl|1\Bigr>$
from the definition of the Glauber--Klauder--Sudarshan coherent
state $|z=1\rangle$.

Calculation can be repeated for an arbitrary coherent state using
generalized Dobi\'nski formulae, \cite{[19]}, for the Bell
polynomials
\begin{equation}
\label{ex18a} B_k(|z|^2)=\sum_{l=0}^kS(k,l)|z|^{2l}=\int_0^\infty\D
x\,x^k\E^{-|z|^2}\sum_{l=0}^{\infty}\frac{|z|^{2l}\delta(x-l)}{l!}\,.
\end{equation}
It leads to
\begin{equation}
\label{ex18aa}
\begin{array}{c}
\disty\Bigl<z\Bigr|\mathcal{N}\Bigl[\exp\left(-\beta(\lambda
a^\dag a+\frac{\lambda^M}{M!}\,(a^\dag
a)^M\right)\biggr)\Bigr]\Bigl|z\Bigr>=
\sum_{n=0}^{\infty}H_n^{(M)}\left(-\beta,\frac{-\beta}{M!}\right)
B_n(|z|^2)\frac{\lambda^n}{n!}=\\[6pt]
\disty\hskip30mm=\E^{-|z|^2}\sum_{k=0}^{\infty}\frac{|z|^{2k}}{k!}\,
\exp\left[-\beta\left(k\lambda+\frac{(k\lambda)^M}{M!}\right)\right],
\end{array}
\end{equation}
As mentioned the series (\ref{ex18aa}) has a physical
interpretation --- for $\beta=1/kT$ it is the \textit{partition
function integrand} for the Hamiltonians related to (\ref{ex11}).
Moreover, the series (\ref{ex18aa}) may be integrated with respect to $|z^2|$ term by term
which leads to the partition function  expressed in terms of the
Jacobi theta functions and their generalizations \cite{[18a]} as
expected. This we consider as a further argument which justifies
the procedure undertaken and suggests that the method will also be
effective in more complicated applications.

The second example is the word ${\hat{w}}=a+a^\dag$ and the normal
ordering problem of
$\mathcal{N}\left(\exp\frac{1}{M!}\,(\lambda\hat{w})^{M}\right)$.
Thus we consider
\begin{equation}
\label{ex21} F(x)=\exp\left(\frac{x^M}{M!}\right),\quad
M=1,\,2,\,3,\,\dots.
\end{equation}
Because
\begin{equation}
\label{ex22} \mathcal{N}\left(\E^{x\hat{w}}\right)=\;
:G_{\hat{w}}(x,a,a^\dag):\;=\;:\E^{x^2/2}\E^{x(a+a^\dag)}:\;,
\end{equation}
we have
\begin{equation}
\label{ex23} V_1^{(\hat{w})}(a,a^\dag)=a+a^\dag\,,\quad
V_2^{(\hat{w})}(a,a^\dag)=1\,,\quad
V_n^{(\hat{w})}(a,a^\dag)=0\quad{\rm for~}\,n>2\,.
\end{equation}
In order to get the expansion of  (\ref{ex22}) we use the modified
Hermite polynomials
\begin{equation}
\label{ex25} h_n(x)=\left(-\frac{\I}{\sqrt{2}}\right)^n
H_n\left(\frac{\I x}{\sqrt{2}}\right),\quad
\exp\left(2x+\frac{x^2}2\right)=
\sum_{n=0}^\infty\frac{h_n(2)}{n!}\,x^n\,.
\end{equation}
In general we get
\begin{equation}
\label{ex26}
\begin{array}{rcl}
Z_M({{L}},{{V}},\lambda)&=&\disty\exp\left(\frac{\lambda^M}{M!}\,
\frac{\D^M}{\D
x^M}\right)\cdot\exp\left(2x+\frac{x^2}2\right)\Bigl|_{x=0}=\\[12pt]
&=&\disty\sum_{n=0}^\infty\frac{h_{Mn}(2)}{n!}
\left(\frac{\lambda^M}{M!}\right)^n,
\end{array}
\end{equation}
from which it is seen that the values of $A_n$ are
$A_{Mn}=\disty\frac{(Mn)!}{(M!)^nn!}\,h_{Mn}(2)$ and zero
otherwise. The particular cases for which the generating functions
are known in closed forms are:
\begin{equation}
\label{ex26-1}
\begin{array}{lcl}
M=1\,,&\quad&A_n=h_n(2)=1,\,2,\,5,\,14,\,43,\,142,\,499,\,1850,\,\dots\,,
\quad n=0,\,1,\,2,\,\dots\,,\\[6pt]
&&\disty
Z_1({{L}},{{V}},\lambda)=\exp\left(2\lambda+\frac{\lambda^2}2\right),
\end{array}
\end{equation}
and
\begin{equation}
\label{ex27}
\begin{array}{lcl}
M=2\,,&\quad&A_{2n}=1,\,5,\,129,\,7485,\,755265,\,116338005,\,\dots\,,\\[6pt]
&&\disty
Z_2({{L}},{{V}},\lambda)=\sum_{n=0}^\infty\frac{h_{2n}(2)}{n!}
\left(\frac{\lambda^2}{2!}\right)^n=
\frac1{(1-\lambda^2)^{1/2}}\exp\left(\frac{2\lambda^2}{1-\lambda^2}\right),
\end{array}
\end{equation}
known as the Doetsch equality. Physically the latter example
corresponds to a (special case of) single mode superfluidity--type
Hamiltonian ${\cal{H}}\sim(a+a^\dag)^2$), \cite{[20]}, while
mathematically it is the solution to the normal ordering of the
exponential of the general $su(1,1)$-Lie algebra element,
\cite{[22]}.

%
\bfg[ht]                     
\bc                         
\resizebox{8.5cm}{!}{\includegraphics{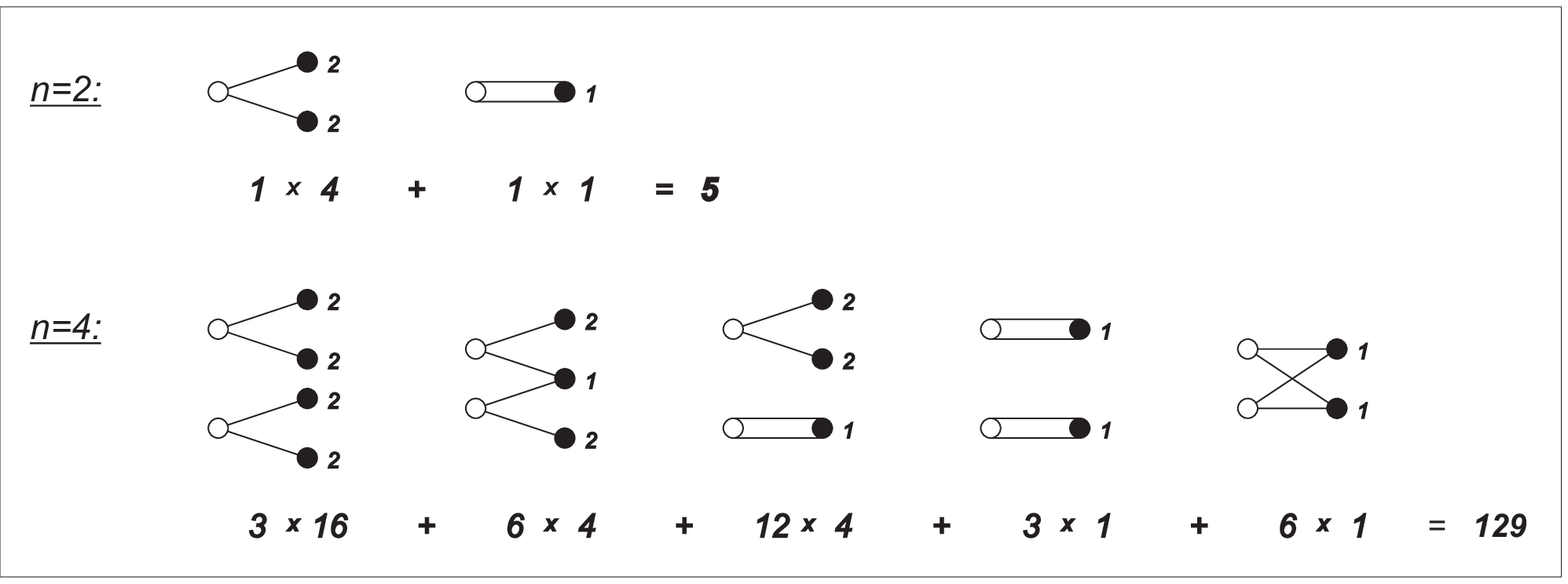}}
\ec                         
\vspace{-7mm} \caption{Lowest order Feynman--type graphs for the
superfluidity--type example with $n=2,4$ lines. The number below
each graph is
$(\mathrm{multiplicity})\times\prod_k(\mathrm{vertex~
factor})V_k=2\delta_{k,1}+\delta_{k,2})$.}
\efg                        

A closed expression has also been found recently, \cite{[23]},
for the case $M=3$:
\begin{equation}
\label{ex28}
\begin{array}{rcl}
Z_3({{L}},{{V}},\lambda)&=&\disty
\sum\limits_{n=0}^\infty\frac{h_{3n}(2)}{n!}
\left(\frac{\lambda^3}{3!}\right)^n=\\[9pt]
&=&\disty\frac1{(1-\phi\lambda^3)^{1/2}}\,
\exp\left(\phi^3\,\frac{\lambda^3}6-\phi^4\,\frac{\lambda^6}8\right)\;
{}_2F_0\left(\frac16,\frac56;-;\frac{3\lambda^6}{2(1-\phi\lambda^3)^{3}}\right),
\end{array}
\end{equation}
where
$\disty\phi(\lambda)=\frac{1-\sqrt{1-4\lambda^3}}{\lambda^3}$ and
${}_2F_0$ is a formal series for (2,0)--generalized hypergeometric
(MacRobert) function. In contrast to the $M=1$ and $M=2$ cases the
series in (\ref{ex28}) is divergent;  however it is asymptotic and
Borel summable to the function of well-defined analytic structure
--- the Airy function \cite{[23-1]}.

\bfg[ht]                     
\bc                         
\resizebox{8.5cm}{!}{\includegraphics{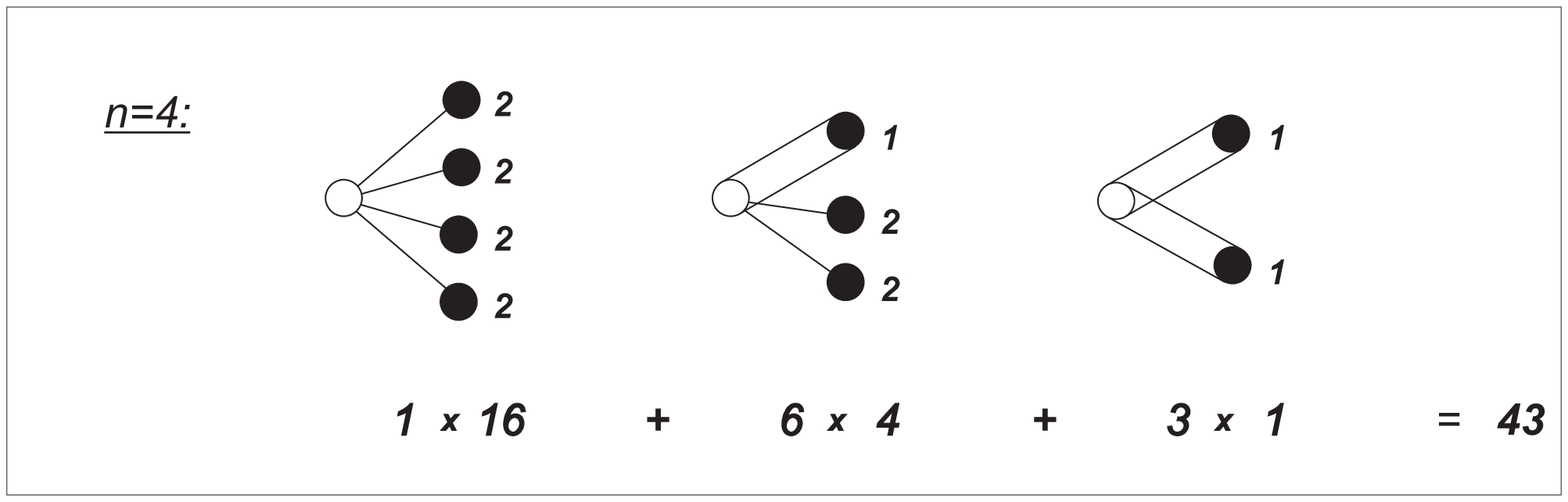}}
\ec                         
\vspace{-7mm} \caption{Feynman--type graphs for the case
$\mathcal{N}\left(\exp\bigl(\frac{1}{3!}\lambda(a+a^{\dagger})^3\bigr)\right)$
with $n=4$ lines.}
\efg                        

The case
$\mathcal{N}\left(\exp\bigl(\frac{y}{4!}(\lambda\hat{w})^{4}\bigr)\right)$
is interesting because it corresponds to the $g\hat{x}^4/4!$
interaction. We demonstrated in the Sect.6 that for its
zero--dimensional analogue the counting formula leads to a
divergent series. Solving the problem according to our scheme we
do not follow the rule (\ref{ex26}) but exploit the knowledge of
$\mathcal{N}\left(\exp\frac{(\lambda\hat{w})^2}{2!}\right)$ given
by Eqn. (\ref{ex27}). Using (\ref{prod4}), (\ref{feynman4-1}),
(\ref{feynman4-2}) and (\ref{ex27}) we have
\begin{equation}
\label{ex29}
\mathcal{N}\left(\exp\left[\frac{y}{4!}\,(\hat{w})^4\right]\right)=
\exp\left(\frac{y}{4!}\,\frac{\D^2}{\D x^2}\right)
\mathcal{N}\bigl(\exp(x\hat{w}^2)\bigr)\Big|_{x=0}
\end{equation}
and subsequently
\begin{equation}
\label{ex30}
\begin{array}{rcl}
\disty \Bigl<1\Bigr|\mathcal{N}\left[\exp\left(\frac{y}{4!}\,
(\lambda\hat{w})^4\right)\right]\Bigl|1\Bigr>&=&\disty
\exp\left(\frac{y}{4!}\,\frac{\D^2}{\D x^2}\right)
\left[\frac1{(1-2x^2)^{1/2}}\,\exp\left(\frac{4x^2}{1-2x^2}\right)\right]
\Big|_{x=0}=\\[9pt]
&=&\disty\sum_{n=0}^\infty\frac{h_{4n}(2)}{n!}
\left(\frac{y\lambda^4}{2(4!)}\right)^n\,.
\end{array}
\end{equation}
The series is divergent for any $y\ne 0$ and according to our best
knowledge up to now no closed form of its generating function has
been found. Investigating its relation to the example of the
Sect.6 we find
\begin{equation}
\label{ex31}
\begin{array}{rcl}
\disty\Bigl<0\Bigr|\mathcal{N}
\left[\exp\left(\frac{y}{4!}\,(\lambda\hat{w})^4\right)\right]\Bigl|0\Bigr>&=&
\disty\lim_{z\rightarrow 0}\Bigl<z\Bigr|\mathcal{N}
\left[\exp\left(\frac{y}{4!}\,(\lambda\hat{w})^4\right)\right]
\Bigl|z\Bigr>=\\[9pt]
&=&\disty \lim_{z\rightarrow 0}\exp\left(\frac{y\lambda^4}{4!}\,
\frac{\D^4}{\D
x^4}\right)\exp\left(2x\mathrm{Re}\,z+\frac{x^2}2\right)\Big|_{x=0}=\\[9pt]
&=&\disty\lim_{z\rightarrow
0}\exp\left(\frac{y\lambda^4}{4!}\,\frac{\D^4}{\D x^4}\right)
\sum_{n=0}^\infty
\frac{x^n}{n!}\,H_n^{(2)}\bigl((2\mathrm{Re}\,z),1\bigr)\Big|_{x=0}=\\[9pt]
&=&\disty
\frac1{\sqrt{\pi}}\sum_{n=0}^{\infty}\frac{\Gamma(2n+\frac12)}{n!}
\left(\frac{2^2y\lambda^4}{4!}\right)^n\,,
\end{array}
\end{equation}
which generalized sum equals to (\ref{feynman6}) in view of
(\ref{feynman4}). A similar technique may be used to find
expansions of analogues of the Green functions and to investigate
their summability by Borel or Pad\'e methods but this important
and interesting topic is out of the scope of this note.

\section{Conclusions}

The main thrust of  this article has been to show the intimate
relationship between the normal ordering problem and
combinatorics, both of series and of graphs.  We have related the
graphical approach to that of  Feynman diagram in  field theory.
Basing our analysis on a simple product formula, we have applied
these combinatorial methods to investigate some simple  problems
of  field theory and many body quantum physics. The close
connection between the normal--ordering problem of the operators
appearing in the theory and the associated combinatorial algebra
is evident. These combinatorial methods open the possibility of
investigating solutions to a variety of field theoretical
problems. Differing from, and more simple than, the standard
methods, this approach encourages us to believe that it may
fruitfully be applied to more complex problems than those which we
have treated for illustration in this article, and that it may
well become a supplement to the standard methods of field theory,
which are at present  dominated by complex and functional
analysis.

\medskip\noindent
{\small We have benefited from the use of the EIS \cite{[25]} in
the course of this work. One of us (PB) wishes to thank the Polish
Ministry of Scientific Research and Information Technology for
support under Grant no: 1P03B 051 26.}

\end{document}